\documentclass[pre,twocolumn,preprintnumbers,amsmath,amssymb]{revtex4}

\usepackage{amsmath,latexsym,epsf,epsfig,graphicx}

\newcommand{\prt}{\partial}
\newcommand{\II}{\mbox{${\mathbb I}$}}

\newcommand{\RR}{\mbox{${\mathbb R}$}}

\newcommand{\DD}{\mbox{${\mathbb D}$}}

\def\UU{\mathbb U}
\def\S{\mathbb S}
\def\T{\mathbb T}
\def\HH{\mathbb H}

\newcommand{\rd}{{\rm d}}

\newcommand{\U}{{\cal U}}

\newcommand{\dQ}{{\dot Q}}
\newcommand{\dS}{{\dot S}}
\newcommand{\W}{{\mathcal W}}

\newcommand{\td}{\widetilde{d}}

\newcommand{\wt}{\hat{t}}

\def\A{\mathcal A} 
\def\H{\mathcal H} 
\def\U{\mathcal U}

\def\M{\mathcal M}

\def\der{\partial }

\def\ri{{\rm i}}

\def\prt{{\partial}}
\def\tr{{\rm Tr}}

\def\Li{{\rm Li}}
\def\e{{\rm e}}

\def\lb{\Omega_{{}_{\rm LB}}}
\def\rlb{\rangle_{{}_{\rm LB}}}
\def\hlb{\H_{{}_{\rm LB}}}

\begin{document}
\title{Quantum Fluctuations of Entropy Production\\  
for Fermionic Systems in the Landauer-B\"uttiker State} 

\author{Mihail Mintchev}
\affiliation{
Istituto Nazionale di Fisica Nucleare and Dipartimento di Fisica dell'Universit\`a di Pisa,\\
Largo Pontecorvo 3, 56127 Pisa, Italy} 

\author{Luca Santoni}
\affiliation{Institute for Theoretical Physics, Princetonplein 5, 3584 CC Utrecht, Netherlands}

\author{Paul Sorba} 
\affiliation 
{LAPTh, Laboratoire d'Annecy-le-Vieux de Physique Th\'eorique, 
CNRS, Universit\'e de Savoie,   
BP 110, 74941 Annecy-le-Vieux Cedex, France}
\bigskip 


\begin{abstract}

The quantum fluctuations of the entropy production for fermionic systems in the 
Landauer-B\"uttiker non-equilibrium steady state are 
investigated. The probability distribution, governing these fluctuations, is explicitly derived 
by means of quantum field theory methods and analysed in the zero frequency limit. 
It turns out that microscopic processes with positive, vanishing and 
negative entropy production occur in the system with non-vanishing probability. In spite of this fact, we show that 
all odd moments (in particular, the mean value of the entropy production) of the 
above distribution are non-negative. This result extends the second principle of thermodynamics to the 
quantum fluctuations of the entropy production in the Landauer-B\"uttiker state. 
The effect of the time reversal is also discussed.

\end{abstract}

\maketitle

\section{Introduction} 
\medskip 

The entropy production is a measure for irreversibility and 
represents an essential characteristic feature of non-equilibrium 
systems. In the quantum context the entropy production is 
fundamental for understanding the deep interplay between 
microscopic and macroscopic physics and in particular, the second 
principle of thermodynamics. For this reason the study of the entropy 
production is receiving a constant attention \cite{SL-78}-\cite{DLNR}. 
A variety of off-equilibrium states \cite{N-07}-\cite{FSU-16} and different 
physical systems \cite{EPR-99}-\cite{DS-17} have been already analysed. In addition, 
the fluctuation relations which have been established \cite{C-99}-\cite{J-11}, 
provide universal information about the nature of the 
entropy production and the related time reversal breaking. 
 
\begin{figure}[h]
\begin{center}
\begin{picture}(600,20)(155,360) 
\includegraphics[scale=0.92]{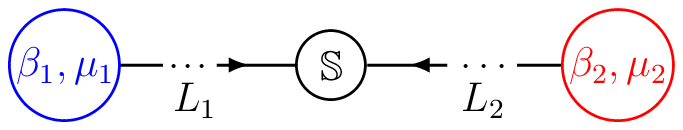}
\end{picture} 
\end{center}
\caption{(Color online) Two-terminal junction.} 
\label{fig1}
\end{figure} 

In this article we investigate the entropy production in {\it quantum} systems 
which are schematically shown in Fig. \ref{fig1}.  Each of the two semi-infinite leads 
$L_i$ is attached at infinity to a heat reservoir $R_i$ with 
(inverse) temperature $\beta_i\geq 0$ and chemical potential $\mu_i\in \RR$. 
The capacity of the reservoirs is assumed to be large enough, so that 
the processes of emission and absorption of particles do not change 
the parameters of $R_i$. A point-like defect is localised at $x=0$ and is described by 
a unitary scattering matrix $\S$. 

The system in Fig. \ref{fig1} models a quantum wire junction \cite{kf-92}-\cite{bcm-09}. The interest 
in such devises, which are essentially one-dimensional systems 
whose transport properties are affected by quantum effects, is largely 
motivated by the fact that they would naturally 
appear in any quantum circuit. Triggered by the remarkable progress 
in nanotechnology, the study of quantum wire junctions nowadays dominates the 
experimental activity in quantum transport. The focus is mainly on the 
particle and heat transport, but recently the entropy production in quantum circuits \cite{camati-16}
and other mesoscopic systems \cite{brun-16} attracts much attention as well. 

The basic physical processes, taking place in the system in Fig. \ref{fig1}, 
can be summarised as follows.  A non-vanishing transmission probability 
$|\S_{12}|^2$ drives the system away from equilibrium, provided that 
the temperatures and/or chemical potentials are different. The departure from 
equilibrium is characterised by the presence of incoming and outgoing 
matter and energy flows from the reservoirs $R_i$. 
The study of these flows started with the pioneering 
work of Landauer \cite{L-57} and B\"uttiker \cite{B-86}, who developed 
an exact scattering approach, going far beyond the linear response 
approximation. The Landauer-B\"uttiker (LB) framework is the basis 
of modern quantum transport theory and has been successfully 
generalised \cite{A-80}-\cite{SI-86} and applied to the computation 
of the noise power \cite{ML-92}-\cite{L-89} and the full counting statistics \cite{LL-92}-\cite{MSS-16}. 

In what follows we apply the LB approach to the study of 
the entropy production. We concentrate on 
fermionic systems, discussing the bosonic case elsewhere \cite{MSS}. 
The basic ingredients of our investigation are: 

\medskip 
(i) a suitably defined field operator $\dS(t,x)$, which describes the 
entropy production;  

(ii) a non-equilibrium steady state $\lb$, which captures the physical 
properties of the system shown in Fig. \ref{fig1}. 

\medskip 
\noindent With this input, all the information about the entropy production is 
codified in the sequence of $n$-point correlation functions ($n=1,2,...$)
\begin{equation}
w_n[\dS](t_1,x_1,...,t_n,x_n) = \langle \dS(t_1,x_1) \cdots \dS(t_n,x_n)  \rlb  \, , 
\label{cf0}
\end{equation} 
the expectation value $\langle \cdots  \rlb$ being computed in the LB state 
$\lb$. 

Previous research in the quantum context has been mainly focussed on 
$w_1[\dS]$, which describes the {\it mean 
value} of the entropy production. Adopting quantum field theory methods, 
we address in this paper the problem of the {\it quantum fluctuations}, which are fully 
characterised by (\ref{cf0}) with $n\geq 2$. The correlation functions $w_n[\dS]$ depend 
on $2n$ space-time variables, which complicate the analysis for large $n$. 
In order to simplify the problem, we follow the standard 
approach \cite{K-87}-\cite{MSS-16} to full counting 
statistics and investigate the {\it zero frequency} limit 
$\W_n[\dS]$ of $w_n[\dS]$, integrating the quantum fluctuations over long period of time. 
We show that in this limit $\W_n[\dS]$ take the form 
\begin{equation} 
\W_n[\dS] = \int_0^\infty \frac{\rd \omega}{2\pi}\, \M_n (\omega)\, , 
\label{mint}
\end{equation}  
where $\omega$ is the energy and $\M_n$ are the {\it moments} of a {\it probability distribution} $\varrho$. 
The derivation of $\varrho$ represents a key point of our 
investigation. In fact, we extract from $\varrho$ the  
basic information about the entropy production at the microscopic level. 
The fundamental quantum process, which takes place in our system, 
is the emission of a particle from the reservoir $R_i$ and the subsequent absorption by $R_j$. 
We derive from $\varrho$ the probability $p_{ij}$ for this  
event at any energy $\omega$ and determine the corresponding entropy production 
\begin{equation} 
\sigma_{ij} = \left [(\beta_i-\beta_j)\omega - (\beta_i\mu_i-\beta_j\mu_j)\right ]|\S_{12}| \, . 
\label{ep1}
\end{equation} 
In the absence of transmission ($\S_{12}=0$) one has $\sigma_{ij}=0$ in agreement 
with the fact that the two heat reservoirs are disconnected and the system is in equilibrium. 
The antisymmetry of $\sigma_{ij}$ implies furthermore that 
the entropy production for emission and absorption of a particle 
by the same reservoir vanishes, as expected on general grounds. 
Moreover, $\sigma_{12}$ and $\sigma_{21}$ have opposite 
sign which, combined with the fact that $p_{12}\not=0$ and $p_{21}\not=0$, leads 
to the conclusion that both processes with positive 
and negative entropy production are necessarily present at the microscopic level. 
Nevertheless, we demonstrate below that the process with positive 
entropy production dominates in the state $\lb$, implying that all moments 
$\{\M_n (\omega)\, :\, \omega \geq 0,\, n=1,2,...\}$ 
of $\varrho$ obey 
\begin{equation}
\M_n (\omega) \geq 0 \, ,
\label{mn}
\end{equation} 
{}for any value of the temperatures and chemical potentials of $R_i$. 
In addition, $\M_n (\omega)$ vanishes for any $\omega$ 
only at the equilibrium $\beta_1=\beta_2$ and $\mu_1=\mu_2$. 

{}For even $n$ the inequality (\ref{mn}) follows directly from the fact that 
$\varrho$ is a true probability distribution on $\RR$, whereas for odd $n$ 
it is a consequence of the specific form of $\varrho$. 
It generalises to the quantum fluctuations the result of Nenciu \cite{N-07} 
\begin{equation}
\langle \dS(t,x) \rlb = \int_0^\infty \frac{\rd \omega}{2\pi}\, \M_1 (\omega) \geq 0   
\label{c1a}
\end{equation}
about the mean value of the entropy production in $\lb$, 
which provides a bridge between microscopic quantum physics and 
the second law of thermodynamics. In this respect, the bound  
(\ref{mn}) represents an extension of the second principle to the quantum 
fluctuations of the entropy production. The result (\ref{mn}) is an intrinsic 
characteristic feature of the LB state. To our knowledge no other steady 
sates with this property are presently 
known. 

The paper is organised as follows. In the next section we describe the 
basic physical properties of the system. We also introduce the entropy 
production operator $\dS$ and the LB representation incorporating 
the non-equilibrium properties of the system in Fig. \ref{fig1}. 
The $n$-point correlation functions of $\dS$ in the LB state 
$\lb$ are derived in section III. In section IV we reconstruct 
the probability distribution $\varrho $ associated with the entropy production, solving 
the corresponding moment problem. The physical properties of $\varrho$ are 
discussed and the role of time reversal is elucidated. It is also shown that 
the presence of a galvanometer in the system does not modify the bound (\ref{mn}). 
Section V is devoted to our conclusions. Finally, the appendices collect 
some technical details.

\section{Preliminaries} 

In this section we summarise the basic non-equilibrium features of 
quantum systems of the type shown in Fig. \ref{fig1}. 
Throughout the paper we adopt the following coordinates 
$\{(x,i)\, ,:\, x\leq 0,\, i=1,2\}$, where $|x|$ denotes 
the distance from the defect and $i$ labels the lead. 

\subsection{Conserved currents and entropy production} 

Let us start by fixing the symmetry content. We consider in this paper physical systems in which  both 
the particle number and the total energy are conserved. Accordingly, the correlation functions are 
invariant under global $U(1)$ transformations and time translations. These symmetries imply 
the existence of a conserved particle and energy currents $(j_t,\, j_x)$ and  $(\theta_{tt},\, \theta_{xt})$. 
Local conservation implies 
\begin{eqnarray}
\der_t j_t(t,x,i) - \der_x j_x(t,x,i) = 0\, ,  
\label{p} \\
\der_t \theta_{tt}(t,x,i) - \der_x \theta_{xt}(t,x,i) = 0\, .
\label{e}
\end{eqnarray} 
In order to generate global conserved charges from $j_t$ and $\theta_{tt}$, which define the 
particle number and total energy respectively, one must impose the Kirchhoff's rules 
\begin{equation} 
\sum_{i=1}^2 j_x(t,0,i) = \sum_{i=1}^2 \theta_{xt}(t,0,i) = 0 \, , 
\label{K}
\end{equation} 
which are assumed in what follows. 

The total energy of our system has two components: heat energy and chemical energy. 
Since the chemical energy density is given by $\mu_ij_t(t,x,i)$, for the heat density 
one has \cite{call} 
\begin{equation} 
q_t(t,x,i) = \theta_{tt}(t,x,i) - \mu_i j_t(t,x,i) \, .  
\label{qd}
\end{equation} 
Accordingly, the heat current reads 
\begin{equation}
q_x(t,x,i) = \theta_{xt}(t,x,i) - \mu_i j_x(t,x,i)\, . 
\label{q}
\end{equation}
{}Following \cite{call}, we introduce at this point the entropy production operator  \cite{JP-01,N-07}
\begin{equation}
\dS (t,x) = - \sum_{i=1}^2 \beta_i\, q_x(t,x,i) \, .  
\label{dS}
\end{equation} 
The definition (\ref{dS}) involves the non-equilibrium heat currents flowing in the leads $L_i$ 
and the equilibrium temperatures $\beta_i$ of the heat reservoirs. The operator 
(\ref{dS}) will be the main subject of our investigation below. 

A simple but deep difference between the heat current $q_x(t,x,i)$ and 
entropy production operator $\dS(t,x)$ is worth stressing. The current $q_x(t,x,i)$ is 
a local observable, which depends on the lead $L_i$ where it 
is observed or measured. The entropy production operator $\dS(t,x)$ 
concerns instead the whole system and does not refer to a single lead. 
Accordingly, the correlation functions (\ref{cf0}), which describe the entropy production fluctuations,  
take into account all the {\it interference effects} between the heat currents in the two different leads 
$L_1$ and $L_2$. The contribution of the interference terms to (\ref{cf0}) 
is fundamental for proving the bound (\ref{mn}). 

It is instructive at this stage to describe the basic physical process taking place  
in the system in Fig. \ref{fig1} and generating the entropy production. 
The conservation laws (\ref{p},\ref{e}) obviously imply the local heat current conservation 
\begin{equation}
\der_t q_t(t,x,i) - \der_x q_x(t,x,i) = 0\, . 
\label{q1}
\end{equation} 
However, if $\mu_1 \not=\mu_2$ the heat current violates the Kirchhoff rule. One has in fact  
\begin{equation} 
\sum_{i=1}^2 q_x(t,0,i) = (\mu_1-\mu_2) j_x(t,0,1) \, .    
\label{Kq}
\end{equation} 
Since the total energy is conserved, both the heat and chemical energies are in general not 
separately conserved. Therefore, for $\mu_1 \not=\mu_2$  the junction in Fig. \ref{fig1} 
operates as {\it energy converter without dissipation} \cite{MSS-14}. The two possible regimes 
are controlled by the expectation value of the operator 
\begin{equation} 
\dQ = - \sum_{i=1}^2 q_x(t,x,i) \,  
\label{qdot}
\end{equation} 
in the underlying non-equilibrium state. If $\langle \dQ\rangle <0$ the 
junction transforms heat to chemical energy. The opposite 
process takes place if instead $\langle \dQ\rangle >0$. A detailed study of 
this phenomenon of energy transmutation in the LB sate $\lb$ has been 
performed in \cite{MSS-14}. 

The above general considerations apply to the system in 
Fig. \ref{fig1} with any dynamics preserving the particle number and total energy. 
In this sense they are universal. For concretely evaluating the quantum fluctuations 
associated with $\dS$, one should fix the dynamics and the non-equilibrium state. 
This is done in the next subsection. 
\bigskip

\subsection{Dynamics and the LB state - the Schr\"odinger junction} 
\medskip 

Non-equilibrium systems of the type in Fig. \ref{fig1} behave in a complicated way and 
the linear response or other approximations are usually not enough for fully describing their complexity.  
For this reason the existence of models, which incorporate the main non-equilibrium 
features, while being sufficiently simple to be analysed exactly, is conceptually very important. 
One such example is provided by particles, which are freely moving along the leads and interact 
only in the junction $x=0$. This hypothesis accounts remarkably well \cite{L-98} for the 
experimental results \cite{K-96} about the noise in mesoscopic conductors 
and has been recently confirmed \cite{FH-17} even in the case of 
fractional charge transport in quantum Hall samples. At the theoretical side, our 
previous analysis in \cite{MSS-14}, \cite{MSS-15}, \cite{MSS-16} and \cite{MSS-17}  
shows that this setup represents an exceptional 
testing ground for exploring general ideas about quantum transport. 

One concrete realisation of the above scenario is the Schr\"odinger junction, where the dynamics along the leads 
is fixed by (the natural units $\hbar =c=k_{\rm B}=1$ are adopted throughout the paper)
\begin{equation}
\left (\ri \prt_t +\frac{1}{2m} \prt_x^2\right )\psi (t,x,i) = 0\, ,  
\label{eqm1}
\end{equation} 
supplemented by the standard equal-time canonical anti-commutation relations.  
The junction represents physically a point-like defect localised at $x=0$.  
The associated interaction determines the scattering matrix $\S$, which  
is fixed by requiring that the bulk Hamiltonian $-\prt_x^2/2m$ admits a 
self-adjoint extension in $x=0$. All such extensions are defined \cite{ks-00}-\cite{k-08} 
by the boundary condition 
\begin{equation} 
\lim_{x\to 0^-}\sum_{j=1}^2 \left [\lambda (\II-\UU)_{ij} +\ri (\II+\UU)_{ij}\prt_x \right ] \psi (t,x,j) = 0\, , 
\label{bc1} 
\end{equation} 
where $\II$ is the identity matrix, 
$\UU$ is a generic $2\times 2$ unitary matrix and $\lambda >0$ is a 
parameter with dimension of mass. Eq. (\ref{bc1}) guaranties unitary time evolution and 
implies \cite{ks-00}-\cite{k-08} the scattering matrix 
\begin{equation} 
\S(k) = 
-\frac{[\lambda (\II - \UU) - k(\II+\UU )]}{[\lambda (\II - \UU) + k(\II+\UU)]} \, ,  
\label{S1}
\end{equation} 
$k$ being the particle momentum. Equation (\ref{S1}) defines a 
meromorphic function in the complex $k$-plane. 

Since {\it scale invariance} preserves the universal features of one-dimensional 
quantum transport \cite{BDV-15} and leads at the same time to relevant simplifications, 
it is instructive to characterise explicitly the scale invariant elements 
in the family (\ref{S1}). For this purpose we first diagonalise ${\UU}$ 
\begin{equation} 
{\cal U}^*\, {\mathbb U}\, {\cal U} = {\mathbb U}_{\rm d}=  
{\rm diag} \left (e^{-2i\alpha_1}, e^{-2i\alpha_2} \right )\, , 
\quad -{\pi\over 2} < \alpha_i \leq {\pi\over 2} \, , 
\label{d11}
\end{equation} 
where $*$ stands for Hermitian conjugation. 
It follows from (\ref{S1}) that the unitary matrix ${\cal U}$ diagonalises 
${\mathbb S}(k)$ for {\it any} $k$ as well. In fact 
\begin{equation} 
{\mathbb S}_{\rm d}(k) = {\cal U}^* {\mathbb S}(k) {\cal U} = \\
{\rm diag} \left ({k+i \eta_1\over 
k-i \eta_1}, {k+i \eta_2\over k-i \eta_2} \right ) \, , 
\label{d3}
\end{equation} 
where 
\begin{equation} 
\eta_i \equiv \lambda\, {\rm tan} (\alpha_i)\, .  
\label{d4}
\end{equation} 
At this point scale invariance implies \cite{bcm-09, CMV-11} the following alternative  
\begin{equation} 
\eta_i = 
\begin{cases} 
0 \quad \; \;  (\alpha_i=0)\, , & \qquad  \text{Neumann b.c.}\, , \\
\infty \quad (\alpha_i=\pi /2)\, , & \qquad  \text{Dirichlet b.c.} \\ 
\end{cases} 
\label{si1}
\end{equation} 
Accordingly, the scale invariant scattering matrices, called also {\it critical points}, are 
$k$-independent and are given by the family 
\begin{equation}
\S = \U\, \S_{\rm d} \, \U^*\, ,  \qquad \U\in U(2)\, , \qquad \S_{\rm d} = {\rm diag}(1,-1)\,   
\label{sinvS}
\end{equation} 
supplemented by the two isolated elements $\S=\pm \II$. The latter   
are not interesting because there is no transmission between the two leads and 
the system is therefore in equilibrium. We adopt  (\ref{sinvS}) in 
section III.A for deriving the mean value $\langle \dS(t,x) \rlb$ at criticality in explicit form.

The scattering states associated to (\ref{S1}) read \cite{M-11} 
\begin{equation}
\chi(k;x)=\left [ \e^{-\ri k x}\, \II +\e^{\ri k x}\, \S^*(k)\right ]\, , \qquad k\geq 0\, , 
\label{ss}
\end{equation} 
Postponing the discussion of the general case, 
let us assume for the moment that $\S(k)$ has no bound states. 
Then, the solution of the quantum boundary value problem 
(\ref{eqm1},\ref{bc1}) is given by
\begin{equation} 
\psi (t,x,i)  = \sum_{j=1}^2 \int_{0}^{\infty} \frac{dk}{2\pi } 
\e^{-\ri \omega (k)t}\, \chi_{ij}(k;x) a_j (k) \, , 
\label{psi1} 
\end{equation} 
where $\omega(k) = {k^2}/{2m}$ is the dispersion relation and the operators 
$\{a_i(k),\, a^*_i(k)\, :\, k\geq 0,\, i=1,2\}$ generate a standard anti-commutation relation algebra 
$\A_+$.

Both (\ref{eqm1}) and (\ref{bc1}) are invariant under global $U(1)$ phase transformations and 
time translations. The relative conserved particle and energy currents have the well known form  
\begin{equation}
j_x(t,x,i)= \frac{\ri }{2m} \left [ \psi^* (\partial_x\psi ) - 
(\partial_x\psi^*)\psi \right ]  (t,x,i)\, ,   
\label{curr1}
\end{equation} 
and 
\begin{eqnarray}
\theta_{xt} (t,x,i) = \frac{1}{4m} [\left (\partial_t \psi^* \right )\left (\partial_x \psi \right ) 
+ \left (\partial_x \psi^* \right )\left (\partial_t \psi \right ) 
\nonumber \\ - 
\left (\partial_t \partial_x \psi^* \right ) \psi - 
\psi^*\left (\partial_t \partial_x \psi \right ) ](t,x,i)\, ,  
\label{en1} 
\end{eqnarray} 
respectively. Plugging the solution (\ref{psi1}) in 
(\ref{curr1},\ref{en1}), one can express the 
heat current (\ref{q}) and therefore the entropy production field operator in terms of the generators 
of $\A_+$. The result is 
\begin{widetext}
\begin{eqnarray}
\dS(t,x)= \frac{\ri}{4m} \int_0^\infty \frac{\rd k}{2\pi} \int_0^\infty \frac{\rd p}{2\pi}\,  
\e^{\ri t [\omega(k) - \omega(p)]} \qquad \qquad \qquad \qquad \qquad 
\nonumber \\ 
\times \sum_{l,j=1}^2 a^*_l(k) \sum_{i=1}^2\beta_i[2\mu_i-\omega(k)-\omega(p)]\Bigl \{
\chi^*_{li}(k;x) \left [\der_{x} \chi_{i j}\right ](p;x) - 
\left [\der_{x} \chi^*_{li}\right ](k;x) \chi_{ij}(p;x) \Bigr \}a_j(p) \, .
\label{ds1}
\end{eqnarray} 
\end{widetext}
This equation defines $\dS(t,x)$ as a quadratic element of the algebra $\A_+$. In order to 
extract the physical information we are interested in, one must fix a representation of $\A_+$.  
The physical setup in Fig. \ref{fig1} suggests to adopt the LB representation of $\A_+$, which  
generalises the equilibrium Gibbs representation to the case of systems driven away from equilibrium 
by a particle and energy exchange with more then one heat reservoir. A field 
theoretical construction of the Hilbert space  
$\{\hlb,\, (\cdot\, ,\, \cdot)\}$ of this representation is given in \cite{M-11}. For deriving the 
expectation values of (\ref{ds1}) one can concentrate on the $2n$-point function 
\begin{eqnarray}
(\lb\, ,\, a^*_{l_1}(k_1) a_{m_1}(p_1)\cdots a^*_{l_n}(k_n) a_{m_n}(p_n)\lb ) \equiv 
\nonumber \\
\langle a^*_{l_1}(k_1) a_{m_1}(p_1)\cdots a^*_{l_n}(k_n) a_{m_n}(p_n)\rlb \, , \qquad 
\label{ncf1}
\end{eqnarray}
which can be represented as a kind of Slater determinant, 
whose explicit form (\ref{cf1}) is given in appendix A. Using (\ref{cf1}) 
we derive in what follows the correlation functions of the operator $\dS$ in the 
LB representation $\hlb$ and discuss the physical implications. 
\bigskip

\section{Entropy production correlation functions} 
\medskip 

\subsection{The one-point function}  
\medskip 

It is natural to start with the one point function $\langle \dS(t,x) \rlb$, which 
gives the mean value of the entropy production in the LB state $\lb$. 
Using (\ref{ds1}) and (\ref{cf1}) for $n=1$, one easily obtains the integral representation (\ref{c1a}) with 
\begin{equation}
\M_1(\omega) = 
\tau(\omega )\, [\gamma_2(\omega)-\gamma_1(\omega)][ d_1(\omega)- d_2(\omega)]\, . 
\label{m1}
\end{equation} 
Here 
\begin{equation}
\tau(\omega ) = 
|\S_{12}(\sqrt{2m\omega})|^2 
\label{tp}
\end{equation} 
is the {\it transmission probability}, 
\begin{equation}
\gamma_i(\omega ) = \beta_i(\omega -\mu_i) \, , \quad i=1,2\,  
\label{gamma}
\end{equation} 
and $d_i(\omega)$ is the Fermi distribution 
\begin{equation}
d_i(\omega)= \frac{1}{1+\e^{\gamma_i(\omega)}}     
\label{fe1}
\end{equation} 
of the reservoir $R_i$. 
One can easily check now that both square brackets $[\cdots ]$ of (\ref{m1}) 
have always the same sign or vanish simultaneously. Therefore, 
\begin{equation}
\M_1(\omega) \geq 0\, ,  
\label{m11}
\end{equation} 
which proves (\ref{mn}) for $n=1$. In addition, $\M_1(\omega)=0$ for 
any $\omega$ implies the equilibrium regime $\beta_1=\beta_2$ and $\mu_1=\mu_2$. 

It is worth mentioning that $\langle \dS(t,x) \rlb$, given by (\ref{c1a}, \ref{m1}), 
is both time and position independent. The $t$-independence follows 
from the energy conservation, whereas the $x$-independence is a consequence of the heat 
current conservation (\ref{q1}). Clearly, these are peculiar properties of the one-point function 
$w_1[\dS]$. The study of $\{w_n[\dS]\, ,:\, n\geq 2\}$ in the next 
subsection reveals a more complicated behaviour. 

Let us explore in conclusion the scale invariant regime. 
At criticality the transmission probability $\tau$ is constant and plugging 
(\ref{m1}) in (\ref{c1a}) one can perform the $\omega$-integration explicitly. The result is 
\begin{widetext}
\begin{equation} 
\langle \dS(t,x) \rlb = (\lambda_2-\lambda_1) \frac{\tau }{2\pi} \left [\frac{1}{\beta_2} \ln \left (1+\e^{\lambda_2}\right ) - 
\frac{1}{\beta_1} \ln \left (1+\e^{\lambda_1}\right )\right ] + 
(\beta_1-\beta_2) \frac{\tau }{2\pi} \left [\frac{1}{\beta_1^2}\, \Li_2 \left (-\e^{\lambda_1}\right ) - 
\frac{1}{\beta_2^2}\, \Li_2 \left (-\e^{\lambda_2}\right )\right ]\, ,
\label{sids}
\end{equation}
\end{widetext}
where $\lambda_i \equiv \beta_i \mu_i$ are dimensionless parameters and $\Li_2$ is the dilogarithm 
function. 

\begin{figure}[ht]
\begin{center}
\includegraphics[scale=0.7]{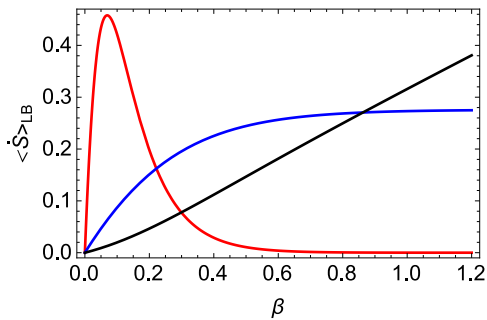}
\end{center}
\caption{(Color online) Entropy production (in temperature units) for $\beta_1=\beta_2 =\beta$ and 
$\tau =1/2$ with $(\mu_1,\mu_2) = (-30,-10)$ (red line),  $(-5,0)$ (blue line) 
and $(4,6)$ (black line).} 
\label{fig2}
\end{figure} 

The mean value the entropy production (\ref{sids}) is generated by both 
the temperature and the chemical potential differences of the heat reservoirs. 
In order to get an idea about the separate effect of these two independent sources, it is 
instructive to consider the limiting regimes $\beta_1=\beta_2$, $\mu_1\not=\mu_2$ 
on one hand and  $\beta_1\not=\beta_2$, $\mu_1=\mu_2$ on the other.  
These ranges of parameters are interesting also from the experimental point of view.  

Let us assume first that that the heat reservoirs have the same temperature 
$\beta_1=\beta_2 = \beta$. 
In this regime the dilogarithms in (\ref{sids}) do not contribute and 
at high temperature one finds 
\begin{equation}
\lim_{\beta \to 0}  \langle \dS(t,x) \rlb^{\beta_1=\beta_2} = 0\, . 
\label{ht}
\end{equation}
The behaviour at low temperature depends on $\mu_i$. Observing that 
$\langle \dS(t,x) \rlb^{\beta_1=\beta_2}$ 
is a symmetric function of $(\mu_1,\mu_2)$, 
one can assume without loss of generality that $\mu_1<\mu_2$ and obtain  
\begin{equation}
\lim_{\beta \to \infty}  \langle \dS(t,x) \rlb^{\beta_1=\beta_2} = 
\begin{cases}
0\, , \quad &{\rm for}\quad \mu_2 < 0\, , \\
\frac{-\mu_1\tau \ln 2}{2\pi} \, , \quad &{\rm for}\quad \mu_2=0\, ,  \\
\infty \, , \quad &{\rm for}\quad \mu_2>0\, ,
\label{nas}
\end{cases}
\end{equation}
as shown in Fig. \ref{fig2}.  

\begin{figure}[ht]
\begin{center}
\includegraphics[scale=0.7]{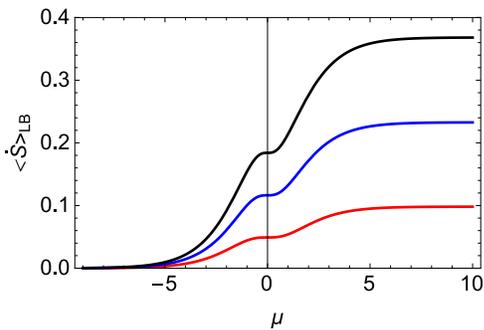}
\end{center}
\caption{(Color online) Entropy production (in temperature units) for 
$\mu_1=\mu_2 =\mu$ and $\tau=1/2$ with $(\beta_1,\beta_2) = (1,2)$ (red line),  $(1,3)$ (blue line) 
and $(1,4)$ (black line).} 
\label{fig3}
\end{figure}

In the second case we set $\mu_1=\mu_2 = \mu$.  The origin of the entropy increase 
is therefore exclusively the difference between the temperatures $\beta_1\not=\beta_2$ 
of the two heat reservoirs. In this case the dilogarithms in (\ref{sids}) 
have a relevant contribution, $\langle \dS(t,x) \rlb^{\mu_1=\mu_2}$ is a symmetric function 
of $(\beta_1,\beta_2)$ and one has 
\begin{equation} 
\begin{split}
\lim_{\mu \to -\infty} \langle \dS(t,x) \rlb^{\mu_1=\mu_2} = 0\, , \qquad \qquad \qquad \qquad \; \; \;  \\
\lim_{\mu \to \infty} \langle \dS(t,x) \rlb^{\mu_1=\mu_2} = \frac{\pi (\beta_1-\beta_2)^2(\beta_1+\beta_2)\tau}
{12 \beta_1^2 \beta_2^2}\, ,  
\end{split}
\label{limit}
\end{equation} 
as displayed in Fig. \ref{fig3}. 

Finally, for $\tau=1$ the defect at $x=0$ is absent and one obtains from (\ref{sids}) the mean entropy 
production of two heat reservoirs connected with a homogeneous infinite lead.

\bigskip 
\subsection{The $n$-point function} 
\medskip 

{}First of all we observe that the correlation function $w_n[\dS]$ depends  
on the time differences 
\begin{equation}
\wt_k \equiv t_k - t_{k+1}\, , \quad \, k=1,...,n-1\, , 
\label{td}
\end{equation}
which is a consequence of the time translation invariance of $\lb$. Since the defect 
at $x=0$ violates translation invariance in space, 
$w_n[\dS]$ depends on all the coordinates $\{x_l\, :\, l=1,...,n\}$ separately.  
In order to simplify the analysis and avoid those variables, which are marginal 
for the entropy production, we introduce the Fourier transforms 
\begin{widetext} 
\begin{equation} 
\W_n[\dS](x_1,...,x_n;\nu ) = \int_{-\infty}^{\infty} \rd \wt_1 \cdots   \int_{-\infty}^{\infty} \rd \wt_{n-1} 
\e^{-\ri \nu (\wt_1+\cdots \wt_{n-1})} w_n[\dS](t_1, x_1,...,t_n,x_n)\, , \qquad n\geq 2\, , 
\label{c3}
\end{equation}
\end{widetext} 
and perform the zero-frequency limit 
\begin{equation}
\W_n[\dS] = \lim_{\nu \to 0^+} \W_n[\dS](x_1,...,x_n;\nu ) \, .
\label{c4}
\end{equation} 
This limit has been adopted already 
in the classical studies \cite{ML-92}-\cite{MSS-15} of quantum noise 
produced by the particle current for $n=2$. It has been extended in \cite{GGM-03} 
to the current cumulants with $n>2$ and applied in the framework 
of full counting statistics \cite{K-87}-\cite{MSS-16} as well. The zero frequency regime 
has a well known physical meaning and is mostly explored in experiments. As mentioned in the introduction, 
in the range of low frequencies all quantum fluctuation are integrated over 
long period of time. It is evident from (\ref{c3}) that in the limit $\nu \to 0$ 
this period becomes actually the whole line. We show in appendix B that 
the structure of $w_n[\dS]$ greatly simplifies in this case. In fact, 
using the definition (\ref{ds1}) of $\dS$ and the correlation function (\ref{cf1}), one finds 
\begin{equation}
\W_n[\dS]=\int_0^\infty \frac{\rd \omega}{2\pi} [\gamma_2(\omega)-\gamma_1(\omega)]^n
\DD_n(\omega) \, . 
\label{w0}
\end{equation} 

The basic steps in deriving the representation (\ref{w0}), as well as 
the explicit form (\ref{nw2}) of the factor $\DD_n(\omega)$ in the integrand, are given in appendix B. 
$\DD_n(\omega)$ is a sum of determinants, which depend on the scattering matrix (\ref{S1}) 
and the Fermi distribution (\ref{fe1}), in other words on $\tau(\omega)$ and $(\beta_i,\mu_i)$. 
It has been shown in \cite{MSS-17} that the bound states of $\S$, if they exists,  
do not contribute in the limit (\ref{c4}) as well.
Despite of these significant simplifications, at the first sight the integrand of 
(\ref{w0}) for generic $n$ might look still complicated. 
As shown in appendix B however, this is not the case and 
the final expression reads 
\begin{equation} 
\W_n[\dS]=\int_0^\infty \frac{\rd \omega}{2\pi}\, \M_n(\omega)\, , 
\label{w2}
\end{equation}
with  
\begin{eqnarray}
\M_{2k-1} &=& \tau^{k} (\gamma_2-\gamma_1)^{2k-1}\, c_1\, , 
\label{nm1}\\
\M_{2k} &=& \tau^{k} (\gamma_2-\gamma_1)^{2k}\, c_2\, .
\label{nm2}
\end{eqnarray} 
Here $ k=1,2,... $, the $\omega$-dependence of all factors has been suppressed for conciseness 
and the following combinations 
\begin{equation}
c_1 \equiv d_1 - d_2 \, , \quad 
c_2 \equiv d_1 + d_2 -2 d_1 d_2\, , 
\label{c}
\end{equation}
have been introduced for convenience. 

The explicit form (\ref{nm1},\ref{nm2}) of the integrands $\M_n$ 
represents a key point of our analysis of the fluctuations of the entropy production. 
First of all, from (\ref{nm1},\ref{nm2}) one infers 
the result (\ref{mn}) announced in the introduction, namely that all $\M_n$ are nonnegative. 
In fact, the argument about the positivity of $\M_1$ applies actually 
for all odd values of $n$. The inequality (\ref{nm2}) for even values of $n$ follows instead from 
\begin{equation} 
c_2 = 
\frac{\e^{\gamma_1} + \e^{\gamma_2}}
{\left (1+\e^{\gamma_1} \right ) \left (1+\e^{\gamma_2} \right )} \geq 0\, .  
\label{mnpeven}
\end{equation} 
Our goal in the next section will be to show that the integrands (\ref{nm1},\ref{nm2}) 
represent indeed the moments of a probability distribution and to reconstruct this distribution.

\bigskip 
 
\section{Probability distribution governing the entropy production} 
\medskip 

The fluctuations of a quantum observable give rise in general to a {\it quasi}-probability distribution. 
Familiar examples are the Wigner function \cite{W-32}, some distributions stemming 
from coherent states in quantum optics \cite{CG-69,F-02} and more recent examples 
associated with time-integrated observables \cite{HC-16, H-17} in the context of full quantum statistics 
\cite{K-87}-\cite{LC-03}. In this section we show that $\dS$ generates in the LB state $\lb$ a {\it true} 
probability distribution $\varrho$. The idea is to reconstruct $\varrho$ from the moments 
(\ref{nm1},\ref{nm2}), solving the underlying moment problem.

\bigskip 
\bigskip 

\subsection{Solution of the moment problem} 

\medskip 

We are looking for a function $\varrho$ with domain $\cal D$ such that 
\begin{equation} 
\M_n = \int_{\cal D} \rd \sigma \, \sigma^n \varrho (\sigma)\, , \qquad n=0,1,... \, , 
\label{Mom1}
\end{equation} 
where $\M_n$ are given for $n\geq 1$ by (\ref{nm1},\ref{nm2}) and 
\begin{equation}
\M_0 = 1 
\label{Mom2}  
\end{equation} 
provides a normalisation condition. The parameter $\sigma$ describes the entropy production. 
There exist \cite{ST-70} three possible choices for the domain $\cal D$ of $\sigma$: the whole line ${\cal D}=\RR$, 
the half line ${\cal D}= \RR_+$ and a compact interval ${\cal D}=[a,b]$. In order to determine $\cal D$ 
we have to investigate the Hankel determinants  
\begin{eqnarray} 
\HH_n \equiv 
\left | \begin{array}{cccccccc}
\M_0&\M_1&\cdots &&  \M_n\\
\M_1&\M_2& \cdots & &\M_{n+1}\\
\vdots &\vdots & \vdots&& \vdots \\
\M_n&\M_{n+1}& \cdots & &\M_{2n}\\
\end{array}\right | \, . 
\label{Mom3}\\
\nonumber 
\end{eqnarray} 

A necessary and sufficient condition for the existence of 
$\varrho$ on $\RR$ is \cite{ST-70} 
\begin{equation}
\HH_n \geq 0 \, , \qquad \forall \, n=1,2,... 
\label{Mom4}
\end{equation} 
Using (\ref{nm1},\ref{nm2},\ref{Mom2}) one gets 
\begin{eqnarray}
\HH_0 = 1\, , \qquad \HH_1= \tau (\gamma_1-\gamma_2)^2(c_2-\tau c_1^2 )\, , \qquad \qquad \;  \; \; 
\label{Mom5}\\
\HH_2 = \tau^3 (\gamma_1-\gamma_2)^6(1-c_2)(c^2_2-\tau c_1^2 )\, , \quad \HH_{n\geq 3}=0\, . \quad 
\label{Mom6}
\end{eqnarray}
Combining the inequalities 
\begin{equation} 
0\leq c_2\leq 1\, , \quad c_2^2-c_1^2 \geq 0\, , 
\label{Mom7}
\end{equation}
which follow directly from the explicit form (\ref{c}) of $c_i$ and using $0\leq \tau\leq1$, 
one gets that both $\HH_1$ and $\HH_2$ are non-negative. Since in addition, 
\begin{equation} 
\HH_2^\prime \equiv 
\left | \begin{array}{cccccccc}
\M_1&\M_2\\
\M_2&\M_3\\
\end{array}\right | = \tau^2 (\gamma_1-\gamma_2)^4 (c^2_1\tau -c^2_2) \leq 0\, ,  
\label{Mom8}
\end{equation} 
the domains $\RR_+$ and $[a,b]$ are excluded \cite{ST-70}. 

Summarising, the entropy production $\sigma $ in the LB state $\lb$ 
gives rise to the so called Hamburger moment problem ${\cal D}=\RR$. 
Moreover, since $\HH_{n\geq 3} = 0$ the  
general theory \cite{ST-70} implies that $\varrho$ 
is fully localised at three different values of $\sigma$. 

Once the domain $\cal D$ has been determined, the explicit form of the distribution $\varrho$ 
can be recovered \cite{ST-70} by performing the Fourier transform 
\begin{equation} 
\varrho (\sigma) = \int_{-\infty}^\infty \frac{\rd \lambda}{2\pi}\, \e^{-\ri \lambda \sigma}\, \varphi (\lambda ) 
\label{Mom9}
\end{equation} 
of the generating function 
\begin{equation} 
\varphi (\lambda) = \sum_{n=0}^\infty \frac{(\ri \lambda)^n}{n!}\, \M_n\, . 
\label{Mom10}
\end{equation} 
Employing (\ref{nm1},\ref{nm2},\ref{Mom2}) one finds 
\begin{eqnarray}
\varphi (\lambda) = 1+ \ri\, c_1\sqrt{\tau}\, \sin \left [\lambda (\gamma_2-\gamma_1)\sqrt{\tau }\, \right ]+ 
\; \; \; \nonumber \\
c_2 \left \{\cos \left [\lambda (\gamma_2-\gamma_1) \sqrt{\tau}\, \right ]-1\right \}\, , 
\label{Mom11}
\end{eqnarray} 
whose Fourier transform reads 
\begin{widetext}
\begin{equation}
\varrho (\sigma ) = \frac{1}{2}(c_2-c_1\sqrt {\tau})\delta\left [\sigma -(\gamma_1-\gamma_2) \sqrt{\tau}\, \right ]
+(1-c_2) \delta(\sigma) + 
\frac{1}{2}(c_2+c_1\sqrt {\tau})\delta \left [\sigma - (\gamma_2-\gamma_1) \sqrt{\tau}\, \right ] \, . 
\label{Mom12}
\end{equation} 
\end{widetext}
Equation (\ref{Mom12}) confirms that the entropy production is 
indeed localised in three points on the $\sigma$-line. It is convenient to adopt at this stage  
the variables $\sigma_{ij}$ defined by (\ref{ep1}), which read   
\begin{equation}
\sigma_{ij} = (\gamma_i-\gamma_j) \sqrt \tau  
\label{Mom13}
\end{equation} 
in terms of $\gamma_i$ and $\tau$. Then $\varrho $ can be rewritten the form 
\begin{equation} 
\varrho (\sigma ) = p_{12}\, \delta (\sigma -\sigma_{12})+ p\, \delta(\sigma) + p_{21}\, \delta (\sigma -\sigma_{21})
\label{Mom14}
\end{equation}
with  
\begin{equation} 
p_{12} = \frac{1}{2}(c_2-c_1\sqrt {\tau})\, ,\quad  
p_{21} = \frac{1}{2}(c_2+c_1\sqrt {\tau})\, , \quad 
p = 1-c_2\, . 
\label{Mom15}
\end{equation} 
Here $p_{ij}$ is the probability of emission of a particle by the reservoir $R_i$ and 
absorption by $R_j$, whereas $p$ is the probability for emission and absorption 
by the same reservoir $R_1$ or $R_2$. One can easily show in fact that 
\begin{equation}
p_{12}+p+p_{21} = 1\, , \quad p_{ij}\in [0,1]\, , \quad p \in [0,1]\, ,  
\label{Mom16}
\end{equation}
implying that $\varrho$ is a true probability distribution. 

It is worth stressing that 
the probabilities (\ref{Mom16}) refer to arbitrary but fixed energy $\omega \in [0,\infty)$. 
At this energy the probabilities for $n$-particle emission and absorption with $n\geq 2$ 
vanish because of Pauli's principle. This is not the case for the bosonic junctions 
discussed in \cite{MSS}, where multi-particle emission/absorption processes 
with the same energy are allowed. 

As anticipated in the introduction, we have shown that both processes with 
positive and negative entropy production appear at the quantum level. 
It is quite intuitive that if the transport of a particle from the red to the blue 
reservoir in the isolated system in Fig. \ref{fig1} increases the entropy, the 
opposite process is decreasing it. The crucial point is that 
according to (\ref{Mom15}) both these events have a non-vanishing probability 
and are present {\it without invoking any time reversal operation}. 

\begin{figure}[ht]
\begin{center}
\includegraphics[scale=0.7]{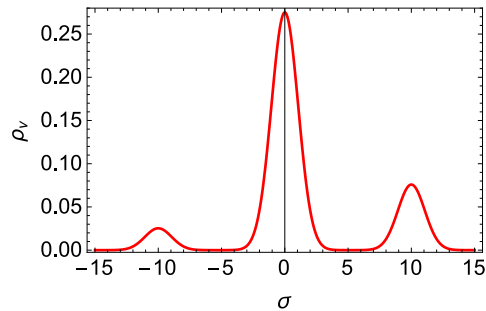}
\end{center}
\caption{(Color online) The smeared distribution $\varrho_\nu$ with $\nu=2/3$, $\gamma_1=21$, $\gamma_2=1$ and $\tau=1/4$.} 
\label{fig4}
\end{figure}

Since $\varrho$ is not smooth but a generalised function, 
in order to illustrate graphically its behaviour it is convenient to introduce 
the $\delta$-sequence 
\begin{equation} 
\delta_\nu(\sigma ) = \frac{\nu}{\sqrt \pi}\, \e^{-\nu^2 \sigma^2}\, , \qquad \nu>0\, , 
\label{delta1}
\end{equation}
and consider the {\it smeared} distribution 
\begin{equation} 
\varrho_\nu (\sigma ) = p_{12}\, \delta_\nu (\sigma -\sigma_{12})+ p\, \delta_\nu (\sigma) + 
p_{21}\, \delta_\nu (\sigma -\sigma_{21})\, . 
\label{delta2}
\end{equation}
As well known, for $\nu \to \infty$ one has $\varrho_\nu \to \varrho$ 
in the sense of generalised functions.  The plots of $\varrho_\nu$ for 
finite values of $\nu$ nicely illustrate the physics behind the distribution $\varrho$. 
One example is reported in Fig. \ref{fig4}. The shape of $\varrho_\nu$ depends  
on $\nu$, but the events with positive entropy production always  
dominate those with negative one. This feature is a consequence of the property 
\begin{equation} 
\sigma_{ij} > 0 \Longrightarrow p_{ij} > p_{ji}\, ,  
\label{delta3} 
\end{equation}
which is $\nu$-independent and holds therefore also in the limit $\nu \to \infty$. 

It is instructive in this point to derive the ratio $P_+/P_-$ where 
$P_\pm$ is the probability to have positive/negative entropy production. 
Without loss of generality one can assume for this purpose that $\sigma_{12}>0$. 
Then 
\begin{equation}
\begin{split}
\frac{P_+}{P_-} &= \frac{p_{12}}{p_{21}}= \frac{c_2-c_1\sqrt {\tau}}{c_2+c_1\sqrt {\tau}} \\
&=\frac{(1-\sqrt {\tau})+(1+\sqrt {\tau})\, \e^{\sigma_{12}/\sqrt {\tau}}}
{(1+\sqrt {\tau})+(1-\sqrt {\tau})\, \e^{\sigma_{12}/\sqrt {\tau}}}\, . 
\end{split}
\label{Mom17}
\end{equation} 
Equation (\ref{Mom17}) generalises the fluctuation relation, discussed in \cite{C-99}-\cite{J-11}, 
to the case in which space translation invariance is broken by a 
quantum point-like defect with transmission probability $\tau$. 
In the limit $\tau \to 1$ the defect disappears, the system becomes 
{\it homogeneous} and one recovers from (\ref{Mom17}) 
the result of Crooks \cite{C-99}  
\begin{equation}
\lim_{\tau \to 1} {\frac{P_+}{P_-}} = \e^{\sigma_{12}}\, , 
\label{Mom18}
\end{equation}
originally obtained in the context of stochastic dynamics. 

Summarising, the probability distribution (\ref{Mom12}) fully describes 
the entropy production zero-frequency fluctuations in 
the LB state $\lb$. It is natural to expect that the behaviour of $\varrho$ depends 
on the choice of this state. This expectation is confirmed in the next subsection, 
where the $\dS$-fluctuations in the state generated from $\lb$ by time reversal 
are explored. 

\bigskip 

\subsection{Impact of time reversal} 
\medskip 

As before, we consider the field $\psi$ defined by (\ref{psi1}) in the 
LB representation $\{\hlb,\, (\cdot\, ,\, \cdot)\}$ of the algebra $\A_+$. 
The time reversal operation acts as usual 
\begin{equation}
T \psi(t,x,i)T^{-1} = \eta_T\, \psi(-t,x,i)\, , 
\label{t1}
\end{equation}
where $|\eta_T|=1$ and $T$ is an anti-unitary operator in $\hlb$ with $T^2=\II$. 
Using (\ref{curr1},\ref{en1}) one easily gets 
\begin{eqnarray}
T\, j_x(t,x,i)\, T^{-1} &=& -j_x(-t,x,i)\, , 
\label{t2}\\
T\,  \theta_{tx}(t,x,i)\, T^{-1} &=&  -\theta_{tx}(-t,x,i)\, , 
\label{t3}
\end{eqnarray}
Since $\langle  j_x(t,x,i) \rlb \not =0$ and 
$\langle  \theta_{tx}(t,x,i) \rlb \not =0$, the overall minus signs in the right hand side 
of (\ref{t2}, \ref{t3}) imply that $T\lb \not = \lb$. Therefore, $T$ generates another state 
$\lb^T = T\lb \in \hlb$ of the system. The entropy fluctuations in this new state are described by 
\begin{equation}
\begin{split}
w_n^T[\dS](t_1,x_1,...,t_n,x_n) = 
\langle \dS(t_1,x_1) \cdots \dS(t_n,x_n)  \rlb^T \\ 
\equiv (T\lb\, ,  \dS(t_1,x_1) \cdots \dS(t_n,x_n)\, T\lb )\, , \qquad
\label{t4}
\end{split}
\end{equation} 
where $(\cdot\, ,\, \cdot)$ is the scalar product in $\hlb$. Using 
(\ref{t2},\ref{t3}) one finds that 
\begin{equation}
w_{2k-1}^T[\dS] = -w_{2k-1}[\dS]\, , \qquad w_{2k}^T[\dS] = w_{2k}[\dS]\, , 
\label{t5}
\end{equation}
with $k=1,2,...$. Therefore, the momenta $\M_n^T$ of the probability distribution 
$\varrho^T (\sigma )$ in the time reversed LB state $\lb^T$ satisfy 
\begin{equation}
\M^T_{2k-1} \leq 0\, , \qquad \M^T_{2k} \geq 0 \, , 
\label{t6}
\end{equation}
which is the mathematical consequence of the physical fact that the processes of emission and absorption 
are inverted with respect to those in $\lb$. 

\bigskip 

\subsection{Comment} 
\medskip 

In the context of particle full counting statistics the possibility to equip the system in Fig. \ref{fig1} with 
a measuring devise, representing a kind of galvanometer, has been also considered in the 
literature \cite{LLL-96}-\cite{ELB-09}. Following 
 \cite{LLL-96}, this alternative scenario can be 
implemented by introducing in (\ref{eqm1}) the minimal coupling $\ri \der_x \longmapsto \ri \der_x + A(x)$ with  
the external field $A(x) \sim \delta(x)$. The physical differences between the two setups have been discussed 
in detail in \cite{LC-03}. Working out the moments of the 
entropy production distribution in the presence of a galvanometer, one finds ($k=1,2,...$)
\begin{eqnarray}
\M^\prime_{2k-1} &=& \tau (\gamma_2-\gamma_1)^{2k-1}\, c_1\, , 
\label{nm1g}\\
\M^\prime_{2k} &=& \tau (\gamma_2-\gamma_1)^{2k}\, c_2\, ,
\label{nm2g}
\end{eqnarray} 
which differ from (\ref{nm1},\ref{nm2}) only by the power of $\tau$. Since $0\leq \tau \leq 1$ 
one concludes that $\M^\prime_n$ satisfy the bound (\ref{mn}) as well. 

The function, generating (\ref{nm1g},\ref{nm2g}), is given by 
\begin{eqnarray}
\varphi^\prime (\lambda) = 1+ \ri\, c_1 \tau\, \sin \left [\lambda (\gamma_2-\gamma_1)\, \right ]+ 
\; \; \; \nonumber \\
c_2 \tau \left \{\cos \left [\lambda (\gamma_2-\gamma_1)\, \right ]-1\right \}\, ,  
\label{Mom11g}
\end{eqnarray} 
and leads to the following probability distribution 
\begin{equation} 
\varrho^\prime (\sigma ) = p^\prime_{12}\, \delta (\sigma -\sigma^\prime_{12})+ p^\prime\, \delta(\sigma) + 
p^\prime_{21}\, \delta (\sigma -\sigma^\prime_{21})\, ,
\label{Mom14g}
\end{equation} 
with  
\begin{equation} 
p^\prime_{12} = \frac{\tau}{2}(c_2-c_1)\, ,\quad  
p^\prime_{21} = \frac{\tau}{2}(c_2+c_1)\, , \quad 
p^\prime = 1-c_2\tau\, ,
\label{Mom15g}
\end{equation} 
and 
\begin{equation}
\sigma^\prime_{ij} = (\gamma_i-\gamma_j)\, . 
\label{Mom13g}
\end{equation} 
One can easily verify that (\ref{Mom15g}) satisfy also in this case (\ref{Mom16}) and define therefore the relative probabilities controlling the particle emission-absorption processes. This feature provides a nice check on the whole setup with a measuring devise. 

In conclusion, the bound (\ref{mn}) is preserved in the presence of a galvanometer as well.

\bigskip 

\section{Outlook and conclusions} 

The present paper pursues further the quantum field theory analysis of the physical properties of the LB non-equilibrium 
steady state. It focuses on the quantum fluctuations of the entropy production in the 
fermionic system shown in Fig. \ref{fig1}. The  junction  acts as a non-dissipative converter of heat 
to chemical potential energy and vice versa.  During the energy transmutation, particles are emitted and 
absorbed by the heat reservoirs, which induces a non-trivial entropy production. Processes with 
positive, vanishing and negative entropy production occur at the quantum level. 
In order to characterise the relative impact of these events, we investigate the correlation 
functions of the entropy production operator in the LB state. The one-point function describes the mean 
entropy production, whereas the $n$-point functions with $n\geq 2$ capture the relative fluctuations. 
We discover that in the zero frequency limit these fluctuations generate a true probability distribution, whose 
moments are all positive. Since the first moment describes the mean entropy production, this remarkable 
property can be interpreted as a kind of extension of the second principle of thermodynamics to the 
non-equilibrium quantum fluctuations in the LB state. The search for other non-equilibrium sates, which share 
the same entropy production properties with the LB state, is a challenging open problem. 

The results of this paper persists even after introducing a galvanometer in the system and 
can be generalised in several directions. Along the above lines 
one can study multi terminal systems as well as the Tomonaga-Luttinger 
liquid away from equilibrium \cite{MS-13,GT-15}. The effect of the quantum statistics 
on the entropy production represents also a deep question, which deserves further study. 
We are currently investigating \cite{MSS} this effect in the bosonic version of the fermion 
system studied above.

\acknowledgments

The work of LS is supported by the Netherlands Organisation for Scientific Research (NWO). 

\appendix

\section{Correlation functions in the LB representation}
\medskip 

In their original work \cite{L-57,B-86} Landauer and B\"uttiker derived the two- and four-point correlation functions 
of $\{a_i(k),\, a^*_i(k)\, :\, k\geq 0,\, i=1,2\}$ in the LB representation $\{\hlb,\, (\cdot\, ,\, \cdot)\}$  
using quantum mechanical tools. If one is 
interested in generic $n$-point functions, it is more convenient to adopt the formalism of 
second quantisation developed in \cite{M-11}. The correlation function (\ref{ncf1}), needed for 
the derivation of the entropy production fluctuations, 
is defined in this formalism by 
\begin{widetext}
\begin{equation}
\langle a^*_{l_1}(k_1) a_{m_1}(p_1)\cdots a^*_{l_n}(k_n) a_{m_n}(p_n)\rangle_{\beta,\mu} = 
\frac{1}{Z} {\rm Tr} \left [\e^{-K} a^*_{l_1}(k_1) a_{m_1}(p_1)\cdots a^*_{l_n}(k_n) a_{m_n}(p_n)\right ]\, , 
\quad k_i>0,\, p_i>0\, , 
\label{A2}
\end{equation} 
where 
\begin{equation}
K= \int_0^\infty \frac{\rd k}{2\pi} \sum_{i=1}^2 \gamma_i[\omega(k)]a^*_i(k) a_i(k) \, , 
\quad Z = {\rm Tr} \left (\e^{-K}\right )\, . 
\label{A3}
\end{equation} 
Referring for the details to \cite{MSS-16, M-11}, we report the final result 
\begin{equation}
\langle a^*_{l_1}(k_1) a_{m_1}(p_1)\cdots a^*_{l_n}(k_n) a_{m_n}(p_n)\rlb = 
\left | \begin{array}{cccccccc}
\Delta_{l_1m_1}(k_1,p_1)&\Delta_{l_1m_2}(k_1,p_2)&\cdots &&  \Delta_{l_1m_n}(k_1,p_n)\\
-{\widetilde{\Delta}}_{l_2m_1}(k_2,p_1)&\Delta_{l_2m_2}(k_2,p_2) & \cdots & &\Delta_{l_2m_n}(k_2,p_n) \\
\vdots &\vdots & \vdots&& \vdots \\
-{\widetilde{\Delta}}_{l_nm_1}(k_n,p_1)&-{\widetilde{\Delta}}_{l_nm_2}(k_n,p_2) & \cdots & &\Delta_{l_nm_n}(k_n,p_n)\\
\end{array}\right | \, . 
\label{cf1}
\end{equation}
\end{widetext} 
Here 
\begin{eqnarray} 
\Delta_{lm}(k,p) &\equiv& 
2\pi \delta (k-p)\delta_{lm}\, d_l[\omega(k)] \, , 
\label{cf2} \\
\widetilde{\Delta}_{lm}(k,p) &\equiv& 
2\pi \delta (k-p)\delta_{lm}\, \td_l[\omega(k)]\, , 
\label{cf3}
\end{eqnarray} 
where $d_l(\omega)$ is the Fermi distribution (\ref{fe1}) and 
\begin{equation}
\td_l(\omega ) = 1-d_l(\omega)=\frac{\e^{\gamma_l (\omega)}}{1+\e^{\gamma_l (\omega)}}\, .  
\label{fe2}
\end{equation}

\bigskip 

\section{Derivation of $\DD_n$}
\medskip 

We summarise first the main steps in deriving the integral representation (\ref{w0}). Using 
(\ref{ds1}) and (\ref{cf1}) one gets a representation of the correlation 
function $w_n[\dS](t_1, x_1,...,t_n,x_n)$ which involves $n$ integrations over $k_i$ and 
$n$ integrations over $p_j$. Then one proceeds as follows: 

(i) by means of the delta functions in (\ref{cf2},\ref{cf3}) one eliminates all $n$ integrals in $p_j$; 

(ii) plugging the obtained expression in (\ref{c3}), one performs all $(n-1)$ integrals in $\wt_l$; 

(iii)  at $\nu=0$ the latter produce $(n-1)$ delta-functions, which allow to eliminate all the integrals 
over $k_i$ except one, for instance that over $k_1=k$;   

(iv) now the curly bracket factor $\{\cdots \}$ in (\ref{ds1}) gives the $x$-independent 
expression 
\begin{eqnarray}
\ri \bigl \{\chi^*_{li}(k;x) \left [\der_{x} \chi_{i j}\right ](k;x) - 
\left [\der_{x} \chi^*_{li}\right ](k;x) \chi_{ij}(k;x) \bigr \} = 
\nonumber \\
-2\ri k [\delta_{li}\delta_{ij} - \S_{li}(k) {\overline \S}_{ji}(k)]\, , \qquad \qquad \qquad 
\label{b0}
\end{eqnarray} 
the bar indicating complex conjugation;

(v) finally, in the integral over $k$ one switches to the variable $\omega =k^2/2m$. 

Following the above steps, one arrives at the integral representation (\ref{w0}) with 
\begin{eqnarray}
\DD_n=
\sum_{i_1,...,i_n=1}^2 
\left | \begin{array}{cccccccc}
\T_{i_1i_1}d_{i_1}&\T_{i_2i_1}d_{i_2}&\cdots &&  \T_{i_ni_1}d_{i_n}\\
-\T_{i_1i_2}\td_{i_1}&\T_{i_2i_2}d_{i_2}& \cdots & &\T_{i_ni_2}d_{i_n}\\
\vdots &\vdots & \vdots&& \vdots \\
-\T_{i_1i_n}\td_{i_1}&-\T_{i_2i_n}\td_{i_2} & \cdots & &\T_{i_ni_n}d_{i_n}\\
\end{array}\right | \, .
\nonumber \\
\label{nw2}
\end{eqnarray} 
Here and to end of this appendix the $\omega$-dependence is omitted for conciseness. The factors  
$d_i$ and $\td_i$ are given by (\ref{fe1}) and (\ref{fe2}) and 
the matrix $\T$, generated by (\ref{b0}), is defined in terms of $\S$ by 
\begin{eqnarray}
\T_{11}&=&-\T_{22} = |\S_{12}|^2 \equiv \tau \, , 
\label{T1} \\
\T_{12}&=&{\overline \T}_{21}=-\S_{11} {\overline\S}_{21}\, . \qquad 
\label{T2}
\end{eqnarray} 

In order to compute $\DD_n$ we introduce 
an auxiliary algebra of fermionic oscillators generated by $\{a_i, a_i^*\, :\, i=1,2\}$, 
which satisfy 
\begin{equation}
[a_i\, ,\, a_j^*]_+ = \delta_{ij}\, ,\qquad 
[a_i\, ,\, a_j]_+ = [a^*_i\, ,\, a_j^*]_+ = 0\, . 
\label{alg}
\end{equation} 
Let us consider the quadratic operators 
\begin{equation}
L= \sum_{i=1}^2 \gamma_i\, a^*_i a_i \, , \qquad 
J= \sum_{i,j=1}^2 a^*_i\, \T_{ij}\, a_j \, . 
\label{b1}
\end{equation}
The key observation now is that $\DD_n$ can be represented in the form 
\begin{equation}
\DD_n = \frac {\tr \left (\e^{-L} J^n \right )}{\tr \left (\e^{-L}\right )}\, , 
\label{b2}
\end{equation}
which can be verified by explicit computation using (\ref{alg},\ref{b1}). One has 
at this point that 
\begin{equation}
\sum_{n=0}^\infty \frac{(\ri \eta)^n}{n!}\, \DD_n =  \frac {\tr \left (\e^{-L}\, \e^{\ri \eta J} \right )}{\tr \left (\e^{-L}\right )}\, . 
\label{b3}
\end{equation}
The right hand side of (\ref{b3}) has been previously computed \cite{MSS-16} for the full counting statistics of the particle 
current (\ref{curr1}). Using the result of \cite{MSS-16}, one finds 
\begin{equation} 
\sum_{n=0}^\infty \frac{(\ri \eta)^n}{n!}\, \DD_n =
1+ \ri c_1 \sqrt{\tau}\, \sin (\eta  \sqrt{\tau}) +c_2 \left [\cos (\eta  \sqrt{\tau})-1\right ] \, , 
\label{b4}
\end{equation}
were $c_i$ are defined by (\ref{c}). From (\ref{b4}) it follows that 
\begin{equation} 
\DD_n  = 
\begin{cases} 
\,1 \, , & n=0\, , \\
\tau^{k}\, c_1 & n=2k-1\, ,\quad \; k=1,2,...\, , \\ 
\tau^{k}\, c_2\, , & n=2k\, ,\quad \qquad k=1,2,...\,  \\ 
\end{cases} 
\label{b5}
\end{equation} 
implying the result (\ref{nm1},\ref{nm2}).

\end{document}